\documentstyle[multicol,epsf,aps,prl,twoside]{revtex} % for two colums
\input entete.tex

\newcommand{\onehalf}{{\scriptstyle {1\over  2}}}
\def\sn{\mathop{\rm sn}\nolimits}

\begin{document}
\draft

\title{\ \\Tests of conformal invariance in randomness-induced
second-order phase transitions}

\author{Christophe Chatelain 
and Bertrand Berche\cite{byline2}}

\address{Laboratoire de Physique des Mat\'eriaux,
Universit\'e Henri Poincar\'e,\\ Nancy
1, Bo\^\i te Postale 239,
F-54506  Vand\oe uvre les Nancy Cedex, France}

\date{Received 6 July 1998; revised manuscript 29 September 1998}

\maketitle

\begin{abstract}
The conformal covariance of correlation functions is checked in the second-order
transition induced by random bonds in the two-dimensional eight-state Potts model.
  The decay of 
correlations  is obtained {\it via} transfer matrix 
calculations in a cylinder geometry,
and large-scale Monte Carlo simulations provide access to
the correlations and the  profiles inside a square with free
or fixed boundary conditions.
In both geometries, conformal transformations constrain the form of the 
spatial dependence, leading to accurate determinations of the order parameter
scaling index,
in good agreement with previous independent determinations obtained through
standard techniques.  The energy density exponent is also computed.
\end{abstract} 

%%%%%%%%%%%%%%%%%%%%%%%%%%%%%%%%%%%%%%%%%%%%%%%%%%%%%%%%%%%%

\begin{multicols}{2}
\narrowtext
It is well known that quenched randomness 
can deeply
affect the critical properties at second-order phase transitions~\cite{harris74},
and is liable to smooth first-order transitions, eventually leading to continuous
transitions~\cite{imrywortis79,aizenmanwehr89huiberker89}.

In random systems, owing to strong inhomogeneities inherent in disorder, the 
usual symmetry properties required by conformal invariance 
(CI)~\cite{belavinpolyakovzamolodchivov84} do not 
hold. However, by averaging
over disorder realizations (denoted by $[\dots]_{av}$),  
translation and rotation 
invariance are restored, and
it is generally believed that conformal invariance techniques should 
apply in principle. Recent results based on 
this assumption have recently been obtained at randomness-induced 
second-order transitions~\cite{picco97,cardyjacobsen97},
but clear evidence of the validity of conformal invariance is still missing.
The question is of both fundamental and practical interest. From the fundamental
point of view, situations are known where a diverging correlation length {\it does
not guarantee} the validity of CI. A few years ago, 
lattice animals were indeed found to be {\it not conformally invariant} although
they display isotropic
critical behavior with correlation lengths satisfying the usual scaling
$\xi\sim L$ with the system size~\cite{millerdebell93}.
If conformal invariance works, on the other hand, its powerful techniques might
be applied with no restriction to investigate the critical 
behavior of 2D random systems~\cite{cardy}.

In the 2D random-bond Ising Model, both analytic~\cite{shalaev94} and numerical 
results~\cite{selkeshchurtalapov94} are available.
Transfer matrix (TM) calculations were also used to study the correlation function
decay along strips and to compute the conformal anomaly (defined as a universal amplitude
in finite-size corrections to the free 
energy)~\cite{dequeirozstinchcombe9297} and, since
disorder is marginally irrelevant in the 2D Ising model, conformal 
invariance techniques
were indeed efficient.
At randomness-induced second-order phase transitions, a direct comparison
between the results deduced from conformal invariance, and standard
techniques, such as finite-size scaling (FSS) Monte Carlo (MC) simulations, have 
nevertheless not yet been made. After the pioneering work of Imry and 
Wortis, the first large-scale MC simulations devoted to the influence
of quenched randomness, in a system whose pure version undergoes a strong
first-order phase transition, were applied to the case of the eight-state
random-bond Potts model 
(RBPM)~\cite{chenferrenberglandau9295}. 

The Hamiltonian of the system with quenched random nearest-neighbor
interactions is written: 
	$-\beta {\cal H}=\sum_{({\bf r},{\bf r'})}K_{{\bf rr'}}
	\delta_{\sigma_{\bf r},\sigma_{\bf r'}}$, 
where the spins, located at sites ${\bf r}$ of a square lattice,
  take the values $\sigma_{\bf r}=1,2,\dots,q$.
The coupling strengths are allowed to take two different values $K$ and $K'=Kr$ 
with probabilities $p$ and $1-p$, respectively.  If both couplings are
distributed with the same 
probability, $p=0.5$, 
the system is, on average, self-dual and the critical point is exactly given 
by duality relations:
$({\rm e}^{K_c}-1)({\rm e}^{K_cr}-1)=q$~\cite{wisemandomany95}. 
This model has again been
carefully investigated recently using both MC 
simulations~\cite{chatelainberche98,picco98} and TM 
calculations~\cite{jacobsencardy97}. The bulk magnetization scaling index
$x_\sigma^b=\beta/\nu$ is $q$ dependent~\cite{jacobsencardy97}. 
Up to now the most refined 
direct estimate of $x_\sigma^b$ in the
case $q=8$ is probably  
a FSS analysis due to Picco ($x_\sigma^b=0.150-0.155$)~\cite{picco98}, 
where it was shown that the random fixed point is 
reached in the range 
$r=8-20$, while crossover effects perturb the results outside this domain.
As mentioned above, TM calculations have already been performed on the 
RBPM~\cite{jacobsencardy97}, but
with a rather weak disorder, $r=2$ leading to $x_\sigma^b=0.142(1)$, 
while standard FSS results are in the range 
$x_\sigma^b=0.158-0.175$~\cite{picco98}. The discrepancy may presumably
be attributed to crossover effects.

%%%%%%%%%%%%%%%%%%%%%%%%%%%%%%%%%%%%%%%%%%%%%%%%%%%%%%%%%%%%%%%%

Due to the increasing literature devoted to
second-order induced phase transitions, in this Rapid Communication, our investigation 
deals with 
the self-dual eight-state RBPM. 
We report results of TM calculations on the strip and
MC simulations in a square geometry, which support the assumption of conformal covariance
of order parameter correlation functions and profiles. We particularly
compare independent determinations of the bulk magnetization exponent, 
resulting in good agreement between standard and conformal 
invariance techniques. Different large-scale simulations are performed in two
types of restricted geometries (strips and squares) with a
ratio $r$ chosen in the range $r=8-20$. The crossover 
regime $r=2$ is also investigated.

In the strip geometry, we used the connectivity transfer matrix formalism of
Bl\"ote and Nightingale~\cite{blotenightingale82}. For a strip of size $L$
with periodic boundary conditions, the 
leading Lyapunov exponent follows from the Furstenberg method:
$\Lambda_0(L)=\lim_{m\to\infty}\frac{1}{m}\ln\left|\!\left|\left(\prod_{k=1}^m 
	{\bf T}_k\right)
	\mid\! v_0\rangle\right|\!\right|$,
where ${\bf T}_k$ is the transfer operator between columns $k-1$ and $k$, and
$\mid\! v_0\rangle$ is a suitable unit initial vector. The leading Lyapunov exponent 
determines the free energy: $f_0(L)=-L^{-1}\Lambda_0(L)$.
For a specific disorder realization, the spin-spin correlation function 
$\langle G_{\sigma}(u)\rangle=\frac{q\langle\delta_{
\sigma_j\sigma_{j+u}}\rangle-1}{q-1}$, where $\langle\dots\rangle$ denotes the 
thermal average, is given by the probability that the spins along some row,  
at columns $j$
and $j+u$, are in the same state:
\begin{equation}
	\langle\delta_{\sigma_j\sigma_{j+u}}\rangle=\frac{
	\langle 0\!\mid{\bf g}_j
	\left(\prod_{k=j}^{j+u-1}
	{\bf T}'_k\right){\bf d}_{j+u}\mid\! 0\rangle}{
	\langle 0\!\mid
	\prod_{k=j}^{j+u-1}
	{\bf T}_k\mid\! 0\rangle},
\label{eq-corr}
\end{equation}
where $\mid\! 0\rangle$ is the ground state eigenvector, ${\bf T}_k'$ is the 
transfer matrix in the extended Hilbert space~\cite{extendedHilbert}, 
${\bf g}_j$ is an operator that identifies the cluster containing $\sigma_j$,
and ${\bf d}_{j+u}$ gives the appropriate weight depending on whether or not
$\sigma_{j+u}$ is in the same state as $\sigma_j$.

It is well known that in disordered spin systems, the strong fluctuations from
sample to sample can induce average 
difficulties~\cite{derrida84}. For that reason we paid 
attention to check, by a careful analysis of the correlation function
probability distribution, that self-averaging problems do not alter the mean
values~\cite{crisantinicolispaladinvulpiani90}. In order to reduce sample
fluctuations, we furthermore considered 
{\it canonical disorder}, a
situation in which exactly the same amount of both couplings is distributed over the
bonds of the system. 
%The stability of the result for the free energy density, compared
%to the standard {\it grand canonical disorder}, is shown in Fig.~\ref{fig-1}
%for different disorder realisations. 
The  computations are then
performed with $10^6$
iterations of the transfer matrix, and the final Lyapunov exponent is
averaged over 20 disorder configurations.

Since our purpose is to check the predictions of conformal symmetry, which are supposed
to be satisfied by {\it average quantities}, i.e., $[\langle{G}_{\sigma}(u)\rangle]_{av}$, our first
aim is to show that, in spite of the lack of self-averaging, our numerical
experiments lead to well-defined
averages. In Ref.~\cite{jacobsencardy97}, Jacobsen and Cardy
argued that in the RBPM, $\ln G$ is self-averaging while $G$ is not. Exploiting
duality, they computed $[{\ln \langle G}_\sigma(u)\rangle]_{av}$ via the free energy of
a system in the presence of a seam of frustrated bonds and a cumulant expansion
enabled them to deduce the behavior of $[\langle{G}_{\sigma}(u)\rangle]_{av}$. In our
approach, we are first interested in the probability distribution of the correlation 
function.%, as shown in Fig.~\ref{fig-2}.
\vskip -0.5cm 
	\begin{figure}
	\epsfxsize=9cm
	\begin{center}
	\mbox{\epsfbox{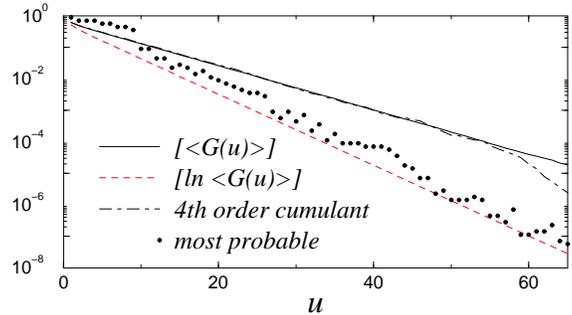}}
	\end{center}%\vskip -0.5cm
	\caption{Average correlation function, most probable value, and
	fourth order 
	cumulant expansion obtained from 63~436 
	replicas for a strip of size $L=6$ ($r=10$).}
	\label{fig-3}  
\end{figure}
\vglue -0.cm 

The most probable value
$G_\sigma^{mp}(u)$ and the average correlation function 
$[\langle{G}_\sigma(u)\rangle]_{av}$, as well as the
averaged logarithm $[{\ln\langle G}_\sigma(u)\rangle]_{av}$, can 
then be deduced at any value
of $u$. Compatible behaviors are found for $G_\sigma^{mp}(u)$ and 
$e^{[{\ln\langle G}_\sigma(u)\rangle]_{av}}$, which confirm the essentially log-normal 
character of the probability
distribution~\cite{crisantinicolispaladinvulpiani90}, in agreement with 
Cardy and Jacobsen. It is thus necessary to perform averages over a
larger number
of samples for $[\langle{G}_\sigma(u)\rangle]_{av}$ than for 
$[{\ln\langle G}_\sigma(u)\rangle]_{av}$ to get the same relative errors.
A cumulant expansion
 enables us to reconstruct $[\langle{G}_\sigma(u)\rangle]_{av}$ and
to compare to the values obtained by averaging directly over the samples. 
The results in Fig.~\ref{fig-3} strengthen the credibility
of the direct average and also clearly show that the 
cumulant expansion up to fourth order still strongly fluctuates compared to 
$[\langle{G}_\sigma(u)\rangle]_{av}$.

\begin{figure}
	\epsfxsize=9cm
	\begin{center}
	\mbox{\epsfbox{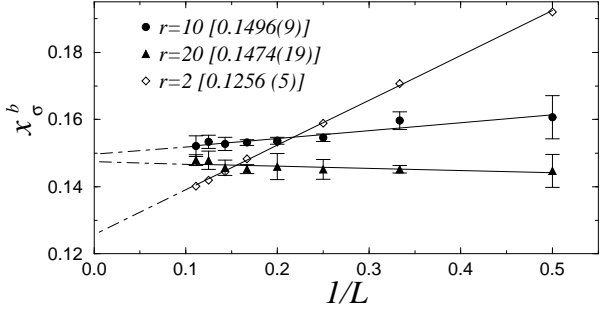}}
	\end{center}%\vskip -0.2cm
	\caption{Magnetic scaling index deduced from the algebraic decay of 
	the average correlation function along
	the strip of size $L$ as a function of $L^{-1}$.}
	\label{fig-4}  
\end{figure}

We will now concentrate on the results that follow from the assumption of
conformal covariance of the averaged correlation functions. In the infinite
complex plane $z=x+{\rm i}y$  at the critical point, the correlation function exhibits the usual
algebraic decay $[\langle{{G}}_\sigma(r)\rangle]_{av}={\rm const}\times r^{-2x_\sigma^b}$, where
$r=\mid\! z\!\mid$. Under a conformal mapping $w(z)$, the correlation functions of
a conformally invariant 2D-system transforms according to
\begin{equation}
	{G}_\sigma(w_1,w_2)=\mid\! w'(z_1)\!\mid^{-x_\sigma^b}
	\mid\! w'(z_2)\!\mid^{-x_\sigma^b} {  G}_\sigma(z_1,z_2).
	\label{eq-transfcorr}
\end{equation}
The logarithmic tranformation $w=\frac{L}{2\pi}\ln z$ is known to map the
$z$ plane onto an infinite strip $w=u+{\rm i}v$ of width $L$ with periodic 
boundary conditions in the transverse direction.
Applying Eq.~(\ref{eq-transfcorr}) in the random system, one gets the usual 
exponential decay along the 
strip $[\langle{G}_\sigma(u)\rangle]_{av}={\rm const}\times
\exp\left(-\frac{2\pi}{L}x_\sigma^b u\right)$, where  the scaling 
index $x_\sigma^b$ can be deduced from a semilog plot. 
For each strip size ($L=2-9$), we realized 40~000 disorder configurations in 
four independent runs, which allowed us to define mean values and error bars. 
The exponent follows from an 
exponential fit in the range 
$u=5--10L$. For $r=10$, the resulting values plotted against $L^{-1}$, 
converge in the $L\to\infty$ limit, towards the final
estimate $x_\sigma^b=0.1496\pm 0.0009$ shown in Fig.~\ref{fig-4}. A
consistent value 
is obtained in the case $r=20$, while in the weak disorder limit the
behavior is drastically different, and strong crossover effects would be expected
for larger strip sizes.
\vskip -0.5cm 
\vglue -0.cm

Another restricted geometry, already used in 
previous FSS MC simulations,
is the square geometry.
The Schwarz-Christoffel mapping  
 $ \zeta=\frac{a}{{2\rm K}}{\rm F}(z,k)$, $z=
  \sn \frac{{2\rm K}\zeta}{a}
  \equiv \sn \left(\frac{{2\rm K}\zeta}{a},k\right)$,
where ${\rm F}(z,k)$ is the elliptic integral of the first kind and $\sn (\zeta,k)$
the Jacobian elliptic sine~\cite{lavrentievchabat},  transforms the upper 
half-plane inside the 
interior of a
square $-a/2\leq{\rm Re}(\zeta)\leq a/2$, 
$0\leq{\rm Im}(\zeta)\leq a$. 
Here, ${\rm K}\equiv{\rm K}(k)$ 
is the complete elliptic integral of the
first kind and the modulus $k$  is a solution of 
${\rm K}(k)/{\rm K}(\sqrt{1-k^2})=\onehalf$.
The correlation function in the semi-infinite geometry is known to take the 
form~\cite{cardy84b}
\begin{equation} 
  {  G}_\sigma(z_1,z)={\rm Cst}.
  (y_1y)^{-x_\sigma^b}\psi(\omega),
  \label{eq: Corhp}
\end{equation}
where the dependence on $\omega=\frac{y_1y}
  {\mid z_1-z\mid^2}$ of the universal scaling function $\psi$ is constrained
by the special conformal transformation. In the random situation
one can again use the transformation
equation~(\ref{eq-transfcorr}) to write the correlations between 
$\zeta_1={\rm i}$, close 
to a side of the square, and
any point inside it, as
follows:
\begin{equation}
  [\langle{  G}_\sigma(\zeta)\rangle]_{av}={\rm const}\times 
  [\mid \zeta'(z)\mid{\rm Im} \left(z(\zeta)\right) ]^{-x_\sigma^b}
  \psi(\omega).
  \label{eq: 10}
\end{equation}
Taking the logarithm of both sides, the bulk critical exponent $x_\sigma^b$ 
can thus be deduced from a  linear fit along $\omega={\rm const}$ 
curves in the square:
\begin{equation}
  \ln [\langle{  G}_\sigma(\zeta)\rangle]_{av}={\rm const'}-x_\sigma^b\ln \kappa(\zeta)+
  \ln\psi(\omega),
  \label{eq: 11}
\end{equation}
with
$\kappa(\zeta)\equiv{{\rm Im}(z(\zeta))} {\left|\left[1-
  z^2(\zeta)\right]\left[1-k^2
  z^2(\zeta)\right]\right|^{-1/2}}.$

\begin{figure}
	\epsfxsize=9cm
	\begin{center}
	\mbox{\epsfbox{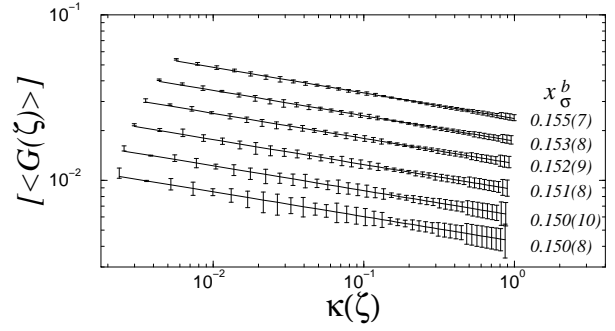}}
	\end{center}%\vskip -0.5cm
	\caption{Rescaled correlation function along six $\omega={\rm const}$ curves in 
	the square ($r=10$).}
	\label{fig-5}  
\end{figure}

The results are shown in Fig.~\ref{fig-5} ($r=10$), in which the Swendsen-Wang 
algorithm has been 
used~\cite{swendsenwang87} for systems of size $101\times 101$ 
and an average was performed over 3000 disorder realizations. Averaging the 
results at different $\omega$'s, one obtains
$x_\sigma^b=0.152\pm 0.003$. Again, a compatible value is  found
at $r=20$ while it disagrees at $r=2$. 
\vskip -0.5cm 
\vglue -0.cm

\begin{figure}
	\epsfxsize=9cm
	\begin{center}
	\mbox{\epsfbox{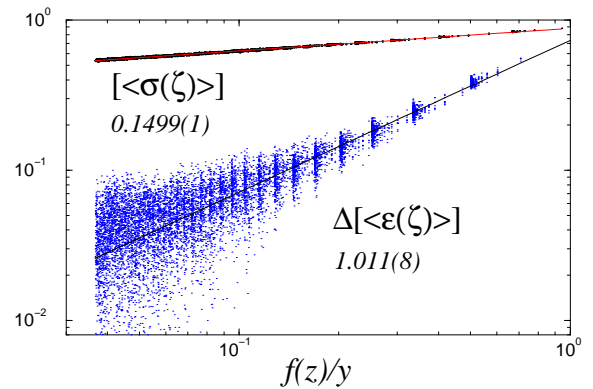}}
	\end{center}\vskip 0.cm
	\caption{Rescaled magnetization and energy density profiles inside the 
	square for 3000 
	disorder realizations ($r=10$). 
	The power law fits are over $100^2$ data points.}
	\label{fig-7}  
\end{figure}

Owing to the unknown scaling function $\psi(\omega)$, the determination is 
not extremely accurate, since a few points are used for the fits. 
It can nevertheless be improved if one considers the magnetization
profile inside a square with fixed boundary conditions. Since it is a one-point function,
its decay from the distance to the surface in the semi-infinite geometry is 
fixed, up to a constant prefactor 
$[\langle\sigma(z)\rangle]_{av}\sim y^{-x_\sigma^b}$.
The local order parameter is 
defined, according to Ref.~\cite{challalandaubinder86}, as the probability for
the spin at site $\zeta$ in the square, to belong to the majority orientation.
After the Schwarz-Christoffel mapping, one gets the following expression
for the average profile in
the square geometry:
\begin{equation}
	[{\langle\sigma (\zeta)\rangle}]_{av}=
	{\rm const}\times\left(\frac{
	\sqrt{|1-z^2(\zeta)|
	|1-k^2z^2(\zeta)|}
	}{
	{\rm Im}(z(\zeta))}\right)^{x^b_\sigma}.
	\label{eq-prof_conf}
\end{equation}
This expression, of the form $[{\langle\sigma (\zeta)\rangle}]_{av}=
[f(z)/y]^{x^b_\sigma}$, holds for any point inside the square. It 
allows an accurate determination of the 
critical exponent 
($x^b_\sigma=0.1499\pm 0.0001$ for $r=10$) via a log-log plot shown in 
Fig.~\ref{fig-7}. We note that in the case $r=2$, the corresponding curve
exhibits a crossover between small and large distances, with  clearly
different exponents close to 0.128 and 0.160, respectively.

\vskip -0.5cm 

 \vbox{
\narrowtext
\begin{table}
\caption{Comparison between FSS and CI determinations of the bulk magnetic
scaling dimension in the $q=8$ RBPM. The quantity that was studied is indicated
in the table as well as the geometry and the numerical technique.
\label{tab1}}
\begin{tabular}{llllll}
\multicolumn{2}{c}{FSS(MC)}&\multicolumn{4}{c}{Conf. Inv.}\\
%\cline{1-2}\cline{3-6}
\multicolumn{2}{c}{square}&strip&strip&\multicolumn{2}{c}{square}\\
%\cline{1-2}\cline{5-6}
SW$^{\rm a}$ & W$^{\rm b}$ & TM$^{\rm c}$ & TM$^{\rm d}$ & \multicolumn{2}{c}{SW}\\
$[\langle M_b\rangle]$ & $[\langle M_b\rangle]$ & $[\langle G(u)\rangle]$ & $[\langle G (u)\rangle]$ 
& $[\langle   G (\zeta)\rangle]^{\rm e}$ & $[\langle \sigma (\zeta)\rangle]^{\rm e}$ \\
\tableline%\tableline
$r=10$  &0.153(1)\\
0.153(3) & 0.152(2) & & 0.1496(9) & 0.152(3) & 0.1499(1) \\
% \tableline
$r=20$ &0.150(1)\\
  & 0.152(2) &  & 0.1474(19) & 0.146(7) & 0.1462(2) \\
%\tableline\tableline
$r=2$ & 0.167(2) & 0.142(1)\\
  & 0.158(3) & 0.142(4) & 0.1256(5) & 0.283(21) & 0.128-0.160 \\
\end{tabular}
\tablenotetext[1]{MC simulations (Swendsen-Wang algorithm, $a\leq 100$, $\sim 500$ samples) from Ref. [14].}
\tablenotetext[2]{MC simulations (Wolff algorithm, $\sim 10^5$ samples) from Ref. [15]. The two values refer
to fits in the ranges $a=20-100$ and 50--200.}
\tablenotetext[3]{TM calculations ($L=1-7$, $10^2$ samples) from Refs. [6,16].}
\tablenotetext[4]{TM calculations ($L=2-9$, $4\times 10^4$ samples), this work.}
\tablenotetext[5]{MC simulations ($a=101$, $3\times 10^3$ samples), this work.}
\end{table}
\narrowtext
}%\vskip -5mm
 
 A summary of our results, compared to independent FSS determinations 
 of the magnetic scaling index, are given in Table~\ref{tab1}. Apart from the
 crossover regime at $r=2$, the agreement is quite good and clearly in favor
 of the validity of the assumption of conformal covariance of correlation 
 functions and profiles {\it at the random fixed point}. 
 On the other hand, in the crossover regime, in which the fixed point is not yet reached, 
 the different techniques lead to different results, and none of them has a
real meaning, since they would presumably be affected by strong crossover 
effects at very large sizes. Since the study of the critical profile with fixed
boundary conditions inside the
square is very accurate, the energy scaling index $x_\varepsilon^b=d-1/\nu$
can also be obtained through the same technique. This dimension has been
calculated in previous studies~\cite{cardyjacobsen97,chatelainberche98} but 
different results were obtained,
possibly contradicting the inequality $x_\varepsilon^b\geq 1$. 
Here, the local energy density
$[\langle\varepsilon (\zeta)\rangle]_{av}$ is 
computed by the average over four bonds on each plaquette. 
It includes a constant bulk 
contribution $[\langle\varepsilon_0\rangle]_{av}$, which is obtained by the
extrapolation to $y\to \infty$ of the profiles
with free and fixed boundary conditions (BC), respectively. At $r=10$, both
BC's yield consistent 
values
$[\langle\varepsilon_0\rangle]_{av}=0.6974$ and $0.6978$. The quantity
$\Delta [\langle\varepsilon (\zeta)\rangle]_{av}=[\langle\varepsilon (\zeta)\rangle]_{av}-0.6976$ 
is then studied as in 
Eq.~(\ref{eq-prof_conf}), and leads to $x_\varepsilon^b=1.011\pm 0.008$ 
(Fig.~\ref{fig-7}). Since the numerical data are very small, there is an important 
dispersion, but the accuracy of the fit remains correct due to the huge number
of data points ($100^2$).
 
 In this Rapid Communication, we have shown that conformal invariance techniques can
 be successfully applied to random systems, and that they lead to
  refined investigations of the critical properties.
  The accuracy, compared to standard techniques, 
 is increased, especially through the magnetization profile inside a square
 where all of the lattice points enter the fit. 
 The critical exponents are finally quite close to the rational values 
 $x_\sigma^b=3/20$ and $x_\varepsilon^b=1$. 
 
  We thank L. Turban for helpful advice, and M. Henkel and J.~L. Cardy 
  for stimulating discussions. The computations were performed at
 the CNUSC in Montpellier under project No. C981009  and the CCH in Nancy.
 The Laboratoire de Physique des Mat\'eriaux is Unit\'e Mixte de Recherche  
 CNRS No. 7556.

%%%%%%%%%%%%%%%%%%%%%%%%%%%%%%%%%%%%%%%%%%%%%%%%%%%%%%%%%%%%%%%%

%%%%%%%%%%%%%%%%%%%%%%%%%%%%%%%%%%%%%%%%%%%%%%%%%%%%%%%%%%%%
\hfil\rule{8pc}{.1mm}\hfil\

\end{multicols}


\begin{references}
\bibitem{extendedHilbert} We used a slightly different technique than Cardy 
	and Jacobsen, whose extended Hilbert space has dimension $Lc_L$, whereas 
	ours has dimension $(L+3)c_L/2$  ($c_L$ is the dimension of the initial
	Hilbert space).
\end{references}
\end{document}